# Qualitative change in structural dynamics of some glass-forming systems

V.N. Novikov[1], A.P. Sokolov[1,2]

[1]*Department of Chemistry and Joint Institute for Neutron Sciences, University of Tennessee, Knoxville, TN 37996, USA*

[2]*Chemical Sciences Division, Oak Ridge National Laboratory, Oak Ridge, TN 37831, USA*

**Abstract**

Analysis of temperature dependence of structural relaxation time $\tau(T)$ in supercooled liquids revealed a qualitatively distinct feature - a sharp, cusp-like maximum in the second derivative of log $\tau_\alpha(T)$ at some $T_{max}$. It suggests that the super-Arrhenius temperature dependence of $\tau_\alpha(T)$ in glass-forming liquids eventually crosses over to an Arrhenius behavior at $T<T_{max}$, and there is no divergence of $\tau_\alpha(T)$ at non-zero $T$. $T_{max}$ can be above or below $T_g$, depending on sensitivity of $\tau(T)$ to change in liquid's density quantified by the exponent $\gamma$ in the scaling $\tau_\alpha(T) \sim \exp(A/T\rho^{-\gamma})$. These results might turn the discussion of the glass transition to the new avenue – the origin of the limiting activation energy for structural relaxation at low $T$.

1. Introduction

The structural relaxation in glass-forming liquids usually shows Arrhenius-like behavior at high temperatures, $\tau_\alpha(T) = \tau_0\exp(E_\infty/T)$, but becomes super- Arrhenius at lower temperatures [1,2]. Moreover, the steepness of the temperature dependence of $\log(\tau_\alpha)$ vs $1/T$ increases sharply with cooling (Fig.1a), meaning that the activation energy for structural relaxation, $E(T)$, increases with decreasing $T$. This suggests that the relaxation time and activation energy might diverge at some finite, non-zero temperature, indicating existence of an underlying phase transition at $T < T_g$ [2]. Attempts to resolve this fundamental question of $\tau_\alpha(T)$ divergence from detailed analysis of experimental data thus far provided different conclusions [3-6]. The authors of [3] found no evidence for the divergence of relaxation time, although they admitted [3]: "*It is not possible to rule out that there is a dynamic divergence of the VFT form, but our findings give no indications of such a divergence*". In ref. [4] it was shown that the divergent signature of $\tau_\alpha$ disappears below $T_g$ in amber. On the other hand, detailed analysis of the relaxation time in poly(vinyl acetate) revealed the VFT-like behavior of $\tau_\alpha$ extends far below $T_g$ (at least by 4 orders) [5,6].

To describe $\tau_\alpha(T)$ various functions were proposed. The most common are three parameter functions: Vogel-Fulcher-Tammann (VFT) function

$$\tau_\alpha = \tau_0\exp(B/(T-T_{VFT})) \text{ [7-9]}; \quad (1)$$

double-Arrhenius [10]

$$\tau_\alpha = \tau_0\exp[(B/T)\exp(E/T)], \quad (2)$$

Bässler-Avramov's [11,12]

$$\tau_\alpha = \tau_0\exp(C/T^\alpha) \quad (3)$$

and parabolic [13]

$$\tau_\alpha = \tau_0\exp[(J/T_0)^2(T_0/T-1)^2] \quad (4)$$

functions. They are based on various phenomenological models, e.g. free-volume [14] and configurational entropy [15], elastic [16], Random First Order Transition (RFOT) [17] and facilitation [18] models, etc. These models either predict the underlying phase transition with diverging relaxation time at finite $T$ (e.g. free volume, entropy based Adam-Gibbs and RFOT), or predict no divergence of $\tau_\alpha(T)$ for any $T$ except at $T = 0$ K. These functions fit the temperature variations of structural relaxation time reasonably well. In some materials they provide good description in the entire temperature range above $T_g$, e.g., VFT function fits $\tau_\alpha(T)$ in polymers or glycerol very well at all $T$. However, they give different predictions on the divergence of $\tau_\alpha(T)$. This divergence would correspond to the divergence of the size of the cooperatively rearranging regions in the Adam-Gibbs approach [15] or of the correlation radius in the Random First Order Theory [17]. Even if there is no divergence of the relaxation time at non-zero $T$, still there is a question does activation energy $E(T)$ diverges as temperature goes to zero (as suggested by e.g. double-Arrhenius equation (2))?

To have deeper understanding of the temperature dependence of $\tau_\alpha(T)$ and to discriminate between various models one should look on more subtle features of the $\tau_\alpha(T)$ behavior. Recent developments in experimental techniques, especially in broadband dielectric spectroscopy, provide highly accurate experimental data that can reveal these subtle changes in $\tau_\alpha(T)$. Here we present new analysis of the temperature dependence of viscosity or $\tau_\alpha$ of supercooled liquids based on their second derivative. We show that at least in some supercooled liquids there is a qualitatively distinct feature in the second derivative of $\tau_\alpha(T)$ that resembles a cusp-like singularity with a sharp maximum. This maximum is not predicted by any of the discussed above 3-parameter functions. Presented analysis suggests that the equilibrium $\tau_\alpha(T)$ turns to Arrhenius-like behavior also at low temperatures, so there is no divergence of $\tau_\alpha(T)$ or $E(T)$ at a finite temperature. The activation energy, in contrast, approach some constant value apparently related to the limited activation energy required for structural relaxation.

## 2. Derivative Analysis

As a first example, we consider the classical glass forming liquid salicylic acid (salol) [19]. The structural relaxation time of salol can be fit reasonably well by several functions discussed above (Fig.1a). The first derivative of $\log\tau_\alpha$ over $T_g/T$ presents the apparent activation energy that increases monotonically with temperature decrease (Fig. 1b). However, the second derivative of the experimental data reveals a sharp peak at a temperature $T_{max} = 255$ K (Fig.2a). A few other independent data for $\tau_\alpha(T)$ of salol [20-22] also reproduce this cusp-like peak in the second derivative. For example, the second derivative of the structural relaxation time of salol measured by a different group [20] (Fig. 2a) exhibits the same peak at the same $T_{max}$ (with accuracy better than 1K). Similar behavior can be found in some other glass-forming systems where the sufficiently accurate data on the relaxation time or viscosity are available [23, 24, 25]. For example, the second derivative of $\log\tau_\alpha(T)$ exhibits the sharp maximum in phenylphthalein dimethyl ether (PDE) and polychlorinated biphenyl with chlorine content 62% (PCB62) (Fig.3); and the second derivative of the $\log\eta$ for the covalent-bonding $B_2O_3$ [25] also exhibits maximum at $T_{max} \sim 630K$ (Fig. 4). However, there are not so many data with accuracy required for the second derivative analysis. Our analysis of large amount of published data revealed that the scattering of the second derivative points is too high to provide any conclusive analysis.

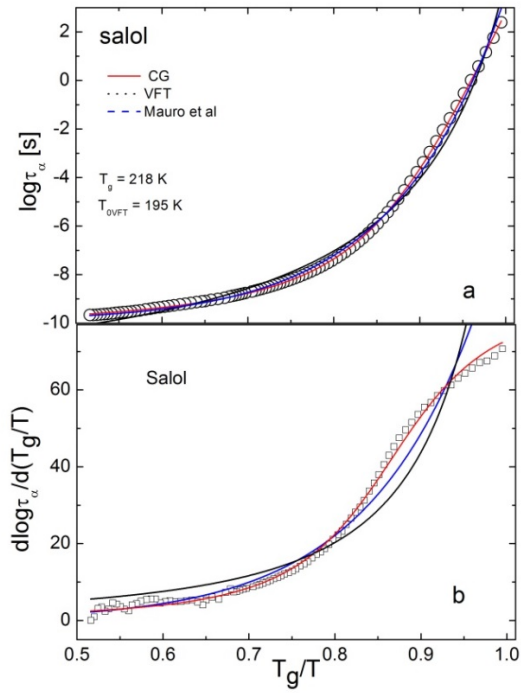

Fig. 1. $\tau_\alpha$ (a) and $d\log\tau/d(T_g/T)$ (b) of salol (symbols). Data for $\tau_\alpha$ are from Ref. [19]. Fits of $\tau_\alpha$ by VFT (dashed line) and Mauro et al [10] (solid line) functions are shown.

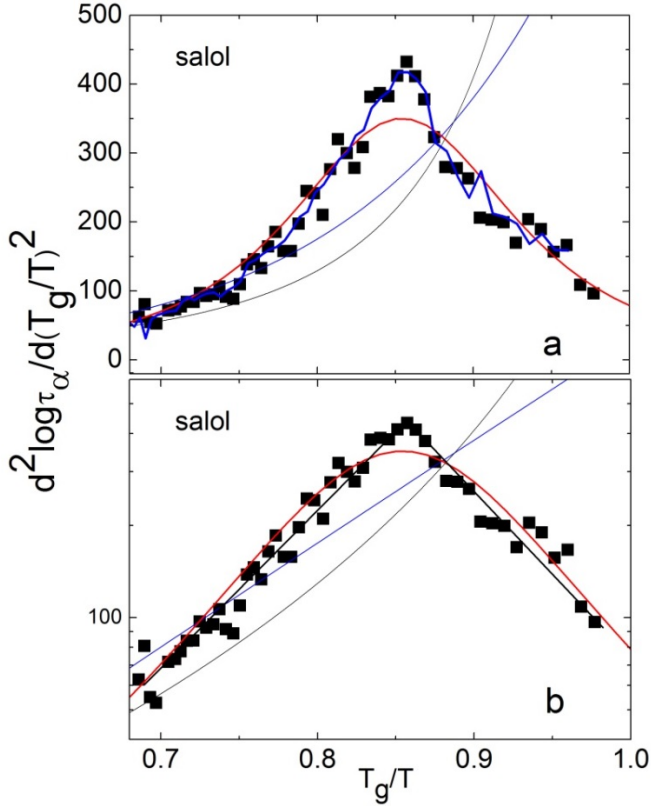

Fig. 2. a). Second derivative of the α-relaxation time over $T_g/T$ in salol (solid squares). Smooth solid red line is the second derivative of the Cohen-Grest function (Eq.(4)) fit of $\log\tau_\alpha$. Dashed blue line is the second derivative of the Mauro et al [10] fitting function, dotted line - the second derivative of the VFT function. The blue solid line is the second derivative of the independent set of data in salol [20].

This analysis also revealed some other glass-forming liquids with sufficiently accurate data that do not exhibit the peak in the second derivative of $\log\tau$ in supercooled state. They include hydrogen-bonding liquids, polymers and room-temperature ionic liquids (RTIL). As examples, we show the second derivative of $\log\tau$ in glycerol and propylene carbonate (PC) (Fig.5), and in tri-cresylphosphat (m-TKP), ethanol, polyvinylacetate (PVAc) and [bmim][NTf2] (Fig.6).

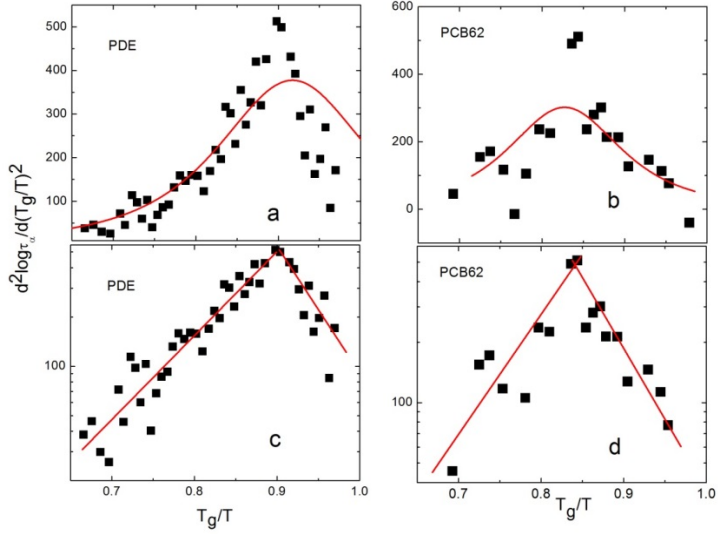

Fig. 3 a). Second derivative of the α-relaxation time $\tau_\alpha$ over $T_g/T$ in phenylphthalein dimethyl ether (PDE) (symbols). Data for $\tau_\alpha$ are from Ref. [23]. Solid line is the second derivative of the Cohen-Grest function fit of $\tau_\alpha$. b). The same for polychlorinated biphenyl with chlorine content 62% (PCB62). Data for $\tau_\alpha$ are from Ref. [24]. c) and d) are the respective data in the logarithmic scale (symbols) and lines present linear approximations (Eq. (2)).

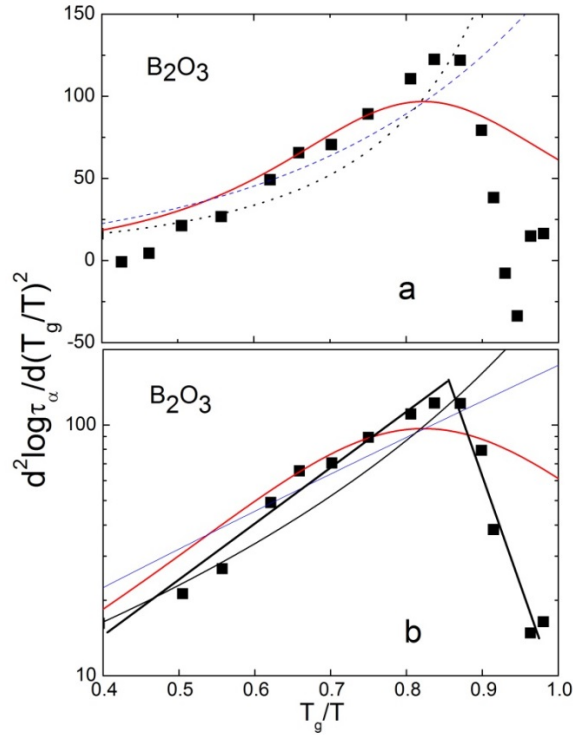

Fig. 4 a). Second derivative of the α-relaxation time over $T_g/T$ in $B_2O_3$ (solid squares). Solid red line is the second derivative of the Cohen-Grest function (Eq.(4)) fit of $\log\tau_\alpha$. Dashed blue line is the second derivative of the Mauro et al [10] fitting function, dotted line - the second derivative of the VFT function. (b) the same plot as (a) but in log scale. The data for $\tau_\alpha$ is from Ref. [25].

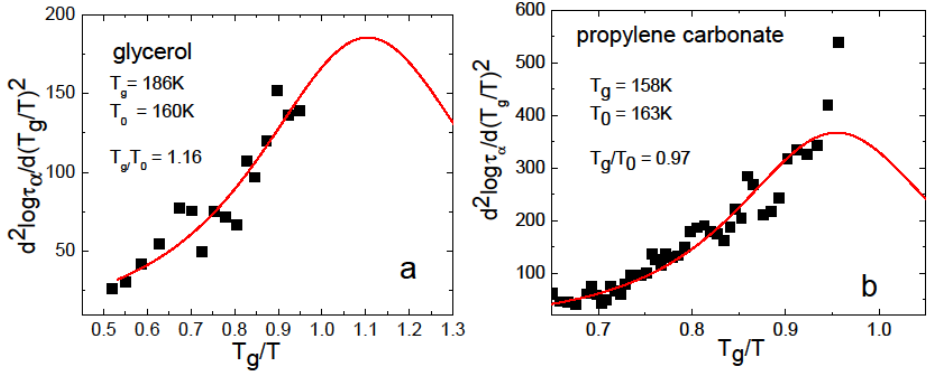

Fig. 5. (a) Second derivative of $\tau_\alpha$ over $T_g/T$ in glycerol (symbols). Data for $\tau_\alpha$ are from Ref. [26]. Solid red line is the second derivative of the Cohen-Grest function (3) that fits $\tau_\alpha$. (b) The same for propylene carbonate, data for $\tau_\alpha$ are from Ref. [23].

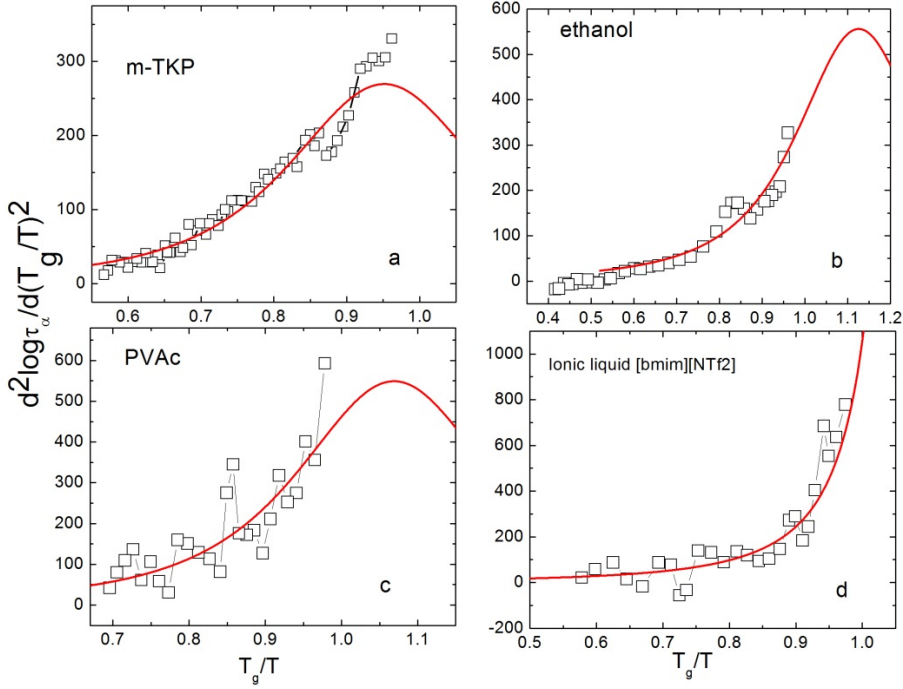

Fig. 6 a) Second derivative of $\tau_\alpha$ over $T_g/T$ in tri-cresylphosphat (m-TKP) (symbols). Data for $\tau_\alpha$ is from Ref. [27]. Solid line presents the second derivative of the Cohen-Grest function that fits $\tau_\alpha$. The same for: b) ethanol (data for $\tau_\alpha$ from Ref. [23]; c) polyvinylacetate (PVAc, data for $\tau_\alpha$ from Ref. [28]); and d) room-temperature ionic liquid [bmim][NTf2] (data for $\tau_\alpha$ from Ref. [29]).

## 3. Discussion

We note, that the second derivative of $\log\tau_\alpha$ over $T_g/T$ is proportional to the first derivative of the apparent activation energy

$$E_a = d\ln\tau_\alpha/d(1/T) \quad (5)$$

The maximum in the second derivative means that the rate with which $E_a$ is growing upon cooling drastically changes behavior at $T_{max}$: The rate increases with decreasing temperature at $T > T_{max}$, while it sharply decreases with further cooling below $T_{max}$. In logarithmic scale, the peak in the second derivative of $\log\tau_\alpha$ can be described by two linear regimes with positive and negative slopes and intersection at $T = T_{max}$ (Figs.2-4). It means that $\log((\log\tau_\alpha)'') = a + b(T_g/T)$ where $a$ and $b$ are some constants, $b > 0$ at $T > T_{max}$ and $b < 0$ at $T < T_{max}$. This corresponds to Arrhenius behavior of $(\log\tau_\alpha)''$ with the activation energy changing sign at $T_{max}$:

$$(\log\tau_\alpha)'' = A_1\exp(E_1/T) \text{ at } T>T_{max} \quad (6a)$$

$$(\log\tau_\alpha)'' = A_2\exp(-E_2/T) \text{ at } T<T_{max}. \quad (6b)$$

For salol, $A_1=7.8*10^{-3}$, $A_2=3.91*10^3$, $E_1 = 2803K$, $E_2 = 2337K$. The apparent activation energy $E_a$, Eq. (5), can be obtained by integrating $(\log\tau)''$:

$$E_a = B_1 + \frac{A_1 T_g^2 \ln 10}{E_1}\exp\left(\frac{E_1}{T}\right), \quad T > T_{max} \quad (7a)$$

$$E_a = E_0 - \frac{A_2 T_g^2 \ln 10}{E_2}\exp\left(-\frac{E_2}{T}\right), \quad T < T_{max} \quad (7b)$$

where $B_1$ and $E_0$ are constants, $B_1 = E_\infty - A_1 T_g^2 \ln 10/E_1 \approx E_\infty$. Eqs.(7a), (7b) predict that there are two Arrhenius regimes: One at high temperatures (with $E_a = E_\infty$ which is well documented [30]), and another one at low temperature ($E_a = E_0$). The activation energy rises with cooling at intermediate temperatures and then saturates at some level. A characteristic temperature interval for the decaying exponential in Eq. (7b) is $\Delta T \sim T_g*(T_g/E_2) \sim 20K$ for salol, i.e. the respective interval is $\Delta(T_g/T) \sim 0.1$. At such distance from $T_{max}$, behavior of $\tau_\alpha(T)$ becomes close to the Arrhenius again. We note that this low-temperature Arrhenius behavior is related to the equilibrium supercooled liquid and is different from the Arrhenius behavior below $T_g$ observed in non-equilibrium glass-formers.

It is important to emphasize that the maximum in the second derivative challenges all the discussed above traditional 3-parameter fitting functions. They produce a monotonic second derivative without any peak (some examples are shown in Fig.3). Thus they all failed to reproduce accurately the temperature variations of $(\log\tau_\alpha)''$ in these liquids even qualitatively in this temperature range. However, there is a four-parameter function derived by Cohen and Grest (CG) [31] in the free-volume percolation model of the glass transition that has the maximum in the second derivative of $\log\tau_\alpha$:

$$\log(\tau_\alpha/\tau_0) = \frac{2B}{T-T_0+\sqrt{(T-T_0)^2+aT}} \quad . \tag{8}$$

Here $T_0$ may be both higher and lower than $T_g$, depending on material. The parameter $a$ is determined by the anharmonicity of the intermolecular potential. It is known that the CG function fits very well the experimental data for $\tau_\alpha(T)$ and $\eta(T)$ in various glass-formers at all $T$ [31,32] . This is not surprising because the CG function has an additional parameter in comparison with the VFT function. The latter is the limiting case of the CG function at $a \rightarrow 0$. The second derivative of the CG function over inverse temperature indeed has a maximum at

$$T_{max} = \frac{T_0}{1-\frac{a}{2T_0}} \quad , \tag{9}$$

although it is not as sharp as the experimental one (Figs. 2-4). Thus, the position of the peak of the second derivative can be determined by simple fitting experimental $\tau_\alpha(T)$ or $\eta(T)$ to the CG function (Eq.(8)). Since the ratio $a/T_0$ is small, ~0.01÷0.1 (Refs. [31,32]), for practical purposes $T_0$ gives a good estimate of $T_{max}$ with accuracy of a few percent.

As it was mentioned in Section 2, some supercooled liquids do not show the peak in the second derivative of $\tau_\alpha(T)$ (Figs.5,6). Fit to the CG function (Eq. (8)) gives $T_0 \sim 160K$ for glycerol which is below its $T_g$ (Fig.5a). This may explain why there is no peak in the second derivative of $\log\tau_\alpha$ in the supercooled glycerol and some other glass-formers: the peak is expected to be at temperatures below $T_g$, where the equilibrium supercooled state cannot be reached experimentally. As one of the consequences, a single VFT or other 3-parameter functions mentioned above can fit $\tau_\alpha(T)$ of glycerol and other materials with $T_0$ below $T_g$ reasonably well in the entire temperature range of supercooled state. This explains the well-known fact that $\tau_\alpha(T)$ in polymers [33], RTIL [29] and some hydrogen-bonding materials [26] can be fit well by a single VFT function, while many molecular liquids require at least two VFT functions, one for low temperatures and another one for high temperatures [19]. We emphasize that the proposed here existence of the maximum in the second derivative of $\tau_\alpha$ at $T_{max}$ (~ $T_0$) below $T_g$ is a speculation based on the fit to the CG function and is not confirmed experimentally. The only justification of this point is that in all cases, when the CG fit provides $T_0 > T_g$ and the data are good enough to analyze the second derivative, there is the maximum at $T_{max} \sim T_0$. It would be important to perform an experiment when a parameter of a glass-former or external conditions, such as pressure, can be varied in order to change the ratio $T_0/T_g$ from $T_0/T_g < 1$ to $T_0/T_g > 1$ and track the evolution of the respective peak of the second derivative of $\log\tau_\alpha$. We note that the CG fit in the case of propylene carbonate estimates $T_0 \sim T_g$ (Fig.5). Although the peak is not resolved (Fig. 5b), the data are consistent with a possible peak at $T \sim T_g$.

The critical question is what controls the position of $T_{max}$ ($\sim T_0$) with respect to $T_g$? The exact physical meaning of the temperature $T_{max}$ is not clear, but in the CG model $T_{max} \sim T_0 = T_1 + a/4 \sim T_1$, where $T_1$ is a parameter showing the sensitivity of the anharmonic part of the inter-particle potential to changing volume [31]. Thus, the stronger anharmonicity of the potential depends on volume, the higher will be $T_{max}$ with respect to some reference material temperature, such as melting or glass transition temperature. Thus, the ratio $T_{max}/T_g$ might correlate with the sensitivity of the structural relaxation time to changing volume. The dependence of the structural relaxation in glass-forming liquids on volume $V$ can be characterized by the exponent $\gamma$ of the so-called thermodynamic scaling [34,35]:

$$\tau(T) = \tau_0 \exp(A/TV^\gamma) \qquad (10)$$

The larger is $\gamma$ the stronger is the dependence of $\tau_\alpha$ on volume. Analysis of $\gamma$ and $T_0$ obtained using CG fit revealed that the ratio $T_0/T_g$ indeed increases with increasing $\gamma$ (Fig. 7). These data suggest that $T_0 > T_g$ in glass-formers with $\gamma \geq 3.5 \div 4$, which are mostly molecular liquids. The peak of the second derivative can be experimentally detected only in such liquids. Materials with $\gamma < 3.5$ (hydrogen-bonding materials, many polymers, RTILs) have $T_0 \leq T_g$. In these materials the peak is predicted to be at temperatures where the supercooled liquid falls out of equilibrium, and thus the peak cannot be observed experimentally.

The presented analysis suggests the following scenario: (i) Glass-forming liquids exhibit Arrhenius-like temperature dependence of the structural relaxation time (viscosity) at high temperatures; (ii) at intermediate temperatures the apparent activation energy for structural relaxation $E_a(T)$ increases upon cooling, and $\tau_\alpha(T)$ exhibits super-Arrhenius behavior; (iii) this increase, however, slows down upon further cooling and (iv) eventually $E_a(T)$ reaches a limiting value, leading to a low-temperature Arrhenius behavior of $\tau_\alpha(T)$ with a constant activation energy $E_0$. Unfortunately, the low-temperature Arrhenius regime in pure form is not observable due to rather long relaxation time required (see e.g. Fig.1b for salol). We want to stress here that this low-temperature Arrhenius is expected in equilibrium supercooled liquid. It should not be confused with the non-equilibrium Arrhenius behavior usually observed at $T < T_g$.

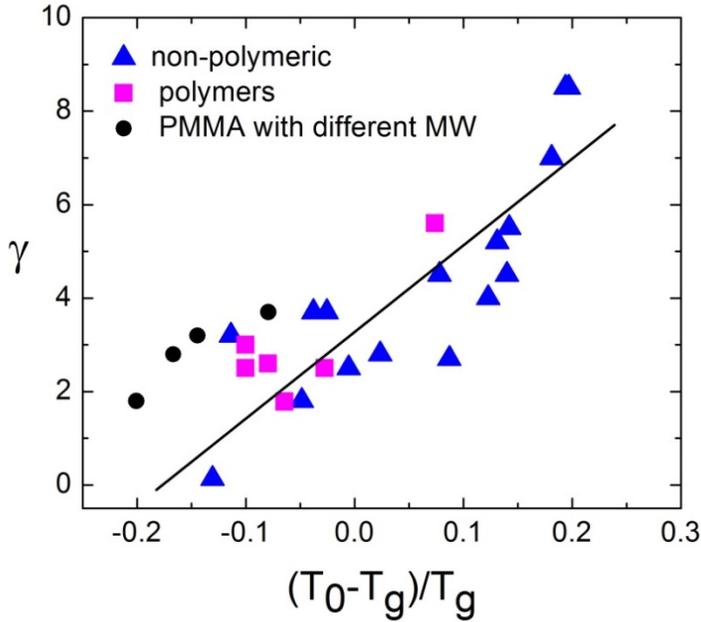

Fig. 7. Correlation between $\gamma$ and $(T_0-T_g)/T_g$. Non polymeric materials (triangles, in increasing $\gamma$ order): sorbitol, glycerol, propylene glycol, 3-fluoroaniline (FAN), diglycidylether of bisphenol A (DGEBA), dibuthylpthalate, propylene carbonate, ortho-terphenyl (OTP), cresolphthalein dimethylether (KDE), phenolphthaleine-dimethyl-ether (PDE), salol, cyclohexane polychlorinated biphenyl (PCB42), 1,1'-bis(p-methoxy phenyl) cyclohexane (BMPC), polychlorinated biphenyl (PCB62), 1,1'-di(4-methoxy-5-methyl phenyl) cyclohexane (BMMPC). Polymers (squares): 1.2 polybutadiene (PB), polystyrene (PS), polypropylene glycol (PPG), polyvinylacetate (PVCa), 1.4 polyisoprene (PI), polymethyl phenyl siloxane (PMPS); polymethyl methacrylate (PMMA) with different molecular weight (circles). The data and references are in the Table 1.

In the Adam-Gibbs [15] and RFOT [17] theories, the activation energy is proportional to the volume of the cooperatively rearranging region (CRR). The crossover to the low-temperature Arrhenius regime means that the size of CRR does not diverge with cooling, and instead, after initial growth eventually saturates at some maximum value. Recently, the low-temperature Arrhenius regime was predicted in a string model [36]. In this model CRR corresponds to strings comprised of fast moving molecules. Applying the theory of living polymers to the strings, the authors showed that the string length increases upon cooling, but will saturate at some limited length at lower temperatures. This would correspond to the limited size of CRR, and consequently, of the activation energy. In elastic models [16] the low-temperature Arrhenius behavior corresponds to the limiting value of shear modulus. In any case, regardless the microscopic mechanism, the activation energy $E$ cannot grow to infinitely large value and will have its limit that depends on the material. Indeed, there should be a limiting energy cost for a molecule to make a relaxation motion in a supercooled liquid. Thus relaxation in any glass-forming liquid eventually will become Arrhenius-like upon cooling and no divergence of time scale at finite $T$ should be expected.

According to Fig. 2, the third order derivative, i.e., the slope of $(\log\tau_\alpha)''$, has a finite jump at $T_{max}$ in salol, and, respectively, the fourth order derivative is infinite at $T_{max}$. In the Adams-Gibbs thermodynamic theory of glass transition $\log\tau_\alpha/\tau_0 = \text{const}/TS_c(T)$ [15] where $S_c(T)$ is the configurational entropy. Thus, $S_c(T)$ should have infinite fourth order derivative at $T_{max}$. This formally means that the system experiences a subtle fourth order phase transition at $T_{max}$. At this point we do not have a clear physical picture of the nature of this transition. We speculate that at decreasing temperature the collective relaxation eventually acquires such high activation energy and CRR size that at $T < T_{max}$ either CRR size is limited by the mechanism of relaxation, as in the string model [36], or another channels of relaxation with limited collectivity have equal or higher rate.

We note that the peak in $B_2O_3$ (Fig.4) looks different from all other cases – it is strongly asymmetric. It is known that $B_2O_3$ exhibits a structural transformation above $T_g$, with increasing number of $B_3O_6$ boroxol rings at the expense of $BO_3$ triangular units [37]. We cannot exclude that the observed maximum in $(\log\eta)''$ in $B_2O_3$ (Fig. 4) is related to this structural change. However, observation of the maximum in the second derivative of several other liquids (Fig.2, 3), and a correlation of $T_0/T_g$ with the scaling parameter γ point to a more general nature of the transition.

The temperature $T_{max}$ at which the increase in $E(T)$ starts to slow down, differs with respect to $T_g$ for different materials and it may be lower or higher than $T_g$ depending on sensitivity of structural relaxation to change in volume (density) (Fig.7). Thus there are systems where crossover to the low-temperature Arrhenius behavior is visible (e.g. salol, PDE, PCB65, $B_2O_3$), but there are systems where this should happen only at $T < T_g$. This explains why attempts to analyze divergence of $\tau_\alpha(T)$ at finite $T$ in various systems [3-6] may produce different results: There are systems (apparently with high γ) where no divergence can be obvious at $T \sim T_g$, while this regime cannot be achieved in other systems, where $T_{max} < T_g$.

## 4. Conclusions

In conclusion, the second derivative of the temperature dependence of the structural relaxation time and viscosity in some supercooled liquids exhibits a sharp maximum. Such a maximum is not predicted by traditional three-parameter functions suggested for description of $\tau_\alpha(T)$. Thus, these functions are missing important qualitative feature of the glass transition. This behavior of the second derivative suggests that the super-Arrhenius dependence of $\tau_\alpha(T)$ should eventually cross over to an Arrhenius regime at further cooling and there is a limiting value for the activation energy required for structural relaxation. The crossover to this low-temperature Arrhenius regime can be both above and below $T_g$, apparently depending on the sensitivity of the structural relaxation of the material to change in volume. This provides a hint to parameters that might define the maximum activation energy for structural relaxation of the liquid. Employing this approach might help to reveal many other peculiarities of dynamics in Soft Matter.

**Acknowledgments:** We acknowledge the support from the NSF Chemistry program (grant CHE-1213444). We are grateful to P. Griffin for useful discussions and bringing to our attention the Cohen-Grest function and to M. Roland for providing experimental data for segmental relaxation in PMMA with different molecular weights.

Table 1. Some parameters of the glass-formers used in the paper.

| | $T_g$, K | $T_0$, K | $\gamma$ | Ref. $\tau_\alpha$ or $\eta$ | Ref. $\gamma$ |
|---|---|---|---|---|---|
| sorbitol | 268 | 233 | 0.16 | 38 | 39 |
| glycerol | 186 | 177 | 1.8 | 40 | 39 |
| 1-propanol | 99 | 96 | 1.89 | 41 | 42 |
| propylene glycol | 168 | 167 | 2.5 | 43 | 44 |
| 3-fluoroaniline (FAN) | 172 | 187 | 2.7 | 45 | 46 |
| dibuthylpthalate | 176 | 156 | 3.2 | 47 | 48 |
| propylene carbonate | 159 | 153 | 3.7 | 43 | 39 |
| OTP | 244 | 274 | 4 | 49 | 39 |
| Cresolphthalein dimethylether (KDE) | 314 | 358 | 4.5 | 23 | 39 |
| Phenolphthaleine-dimethyl-ether (PDE) | 294 | 317 | 4.5 | 23 | 39 |
| salol | 221 | 250 | 5.2 | 19 | 39 |
| polychlorinated biphenyl PCB42 | 225 | 257 | 5.5 | 24 | 39 |
| BMPC 1,1'-bis(p-methoxy phenyl) cyclohexane | 243 | 287 | 39 | 24 | 39 |
| polychlorinated biphenyl PCB62 chlorine content 62% | 274 | 328 | 8.5 | 24 | 39 |
| BMMPC 1,1'-di(4-methoxy-5-methyl phenyl) cyclohexane | 263 | 314 | 8.5 | 24 | 39 |
| [bmim][NTf2] | 181 | 152 | 2.85 | 29 | 50 |
| | | | | | |
| 1,2 polybutadiene (PB) | 253 | 233 | 1.9 | 44 | 44 |
| polystyrene (PS) | 366 | 356 | 2.5 | 51 | 52 |
| polypropylene glycol (PPG) | 202 | 182 | 2.5 | 53 | 39 |
| polyvinylacetate (PVAc) | 302 | 278 | 2.6 | 28 | 39 |
| diglycidylether of bisphenol A (DGEBA) | 254 | 260 | 2.8 | 32 | 39 |
| 1.4 polyisoprene (PI) | 202 | 182 | 3 | 54 | 39 |
| poly(methyl phenyl siloxane) (PMPS) | 243 | 261 | 5.6 | 55 | 39 |
| | | | | | |
| PMMA | 379 | 303 | 1.8 | 56 | 56 |
| PMMA decamer | 288 | 240 | 2.8 | 56 | 56 |
| PMMA tetramer | 240 | 205 | 3.2 | 56 | 56 |
| PMMA trimer | 210 | 193 | 3.7 | 56 | 56 |